\documentclass[final]{aipproc}
\layoutstyle{8x11single}

\usepackage{amssymb,amsmath,amstext,amsthm,amsfonts}

\newcommand{\plotscale}{.73}

\newcommand{\reffig}[1]{Fig.~\ref{#1}}

\newcommand{\refcite}[1]{Ref.~\cite{#1}}
\newcommand{\refscite}[1]{Refs.~\cite{#1}}
\newcommand{\refetal}[1]{\emph{et~al.}~\cite{#1}}
\newcommand{\atom}[2]{\mbox{$^{#1}\text{#2}$}}
\newcommand{\carbon}{{\atom{12}{C}}}
\newcommand{\oxygen}{{\atom{16}{O}}}

\begin{document}

\title{Neutrino induced pion production\\at MiniBooNE and K2K energies}



\author{T.~Leitner}{address={Institut f\"ur Theoretische Physik, Universit\"at Giessen, Germany}}
\author{O.~Buss}{address={Institut f\"ur Theoretische Physik, Universit\"at Giessen, Germany}}
\author{U.~Mosel}{address={Institut f\"ur Theoretische Physik, Universit\"at Giessen, Germany}}
\author{L.~Alvarez-Ruso}{address={Departamento de F\'isica, Centro de F\'isica Computacional, Universidade de Coimbra, Portugal}}

\begin{abstract}
  We investigate charged and neutral current neutrino-induced incoherent pion production
  off nuclei within the GiBUU model at energies relevant for the MiniBooNE and K2K
  experiments.  Special attention is paid to the entanglement between measured CCQE and
  CC$1\pi^+$ cross sections. We further give predictions and compare to recent data
  measured at MiniBooNE.
\end{abstract}

\maketitle

\section{Introduction}

A proper understanding of neutrino induced pion production is essential for the
interpretation of current neutrino oscillations experiments.  A good knowledge of the
neutrino energy is required for a precise determination of oscillation parameters in
$\nu_\mu$ disappearance measurements. These experiments search for a distortion in the
neutrino flux at the detector positioned far away from the source.  By comparing both,
near and far neutrino energy spectra, one gains information about the oscillation
probability and with that about mixing angles and mass squared differences.  However, the
neutrino energy is not measurable directly but has to be reconstructed from the reaction
products. Present oscillation experiments use the CCQE reaction both as signal event and to
reconstruct the neutrino energy from the outgoing muon with two-body kinematics assuming
the target nucleon is at rest. CCQE is defined as $\nu_\ell n \to \ell^- p$ on a single
nucleon; in the nucleus, CCQE is masked by final-state interactions (FSI). Thus, the
correct identification of CCQE events is directly related to the question of how FSI
influence the event selection. The main background to CCQE is CC$1\pi^+$ production. If
the pion is absorbed in the nucleus and/or not seen in the detector, these events can be
misidentified as CCQE. Consequently, a proper understanding of CCQE \emph{and} CC$1\pi^+$
is essential for the reconstruction of the neutrino energy.

The main task in a $\nu_e$ appearance experiment like MiniBooNE is to detect electron
neutrinos in a (almost) pure $\nu_\mu$ beam. The signal event, the $\nu_e$ CCQE
interaction, is dominated by background. A major problem comes from misidentified events,
mainly because of the fact, that the MiniBooNE detector cannot distinguish between a
photon and an electron.  Thus, $\nu_\mu$ induced neutral current $\pi^0$ production, where
the $\pi^0$ decays into two $\gamma$s, is the major source of background when one of the
photons is not seen or both Cherenkov rings overlap. 

As all of the present oscillations experiments use nuclear targets, it is mandatory to
consider FSI, i.e., pion rescattering, with and without charge exchange, and absorption in
the nuclear medium. A realistic treatment of the FSI can be achieved in the framework of a
coupled-channel transport theory --- the GiBUU model. 

After a brief review of our model, we first discuss the impact of pion production on CCQE
measurements. We further investigate the influence of nuclear effects on CC$1\pi^+$ and
NC$1\pi^0$ cross sections, and, where possible, we confront our model to recent data
measured at MiniBooNE.

\section{GiBUU model} 

We treat neutrino-nucleus scattering as a two-step process. In the initial-state step, the
neutrinos interact with nucleons embedded in the nuclear medium.  In the final-state step,
the outgoing particles of the initial reaction are propagated through the nucleus using a
hadronic transport approach.

In the energy region relevant for MiniBooNE, SciBooNE and K2K, the elementary $\nu N$
reaction is dominated by two processes: quasielastic scattering and the excitation of the
$\Delta$ resonance (P$_{33}$(1232)). Additionally, our model includes the excitation of 12
$N^*$ and $\Delta$ resonances with invariant masses less than 2 GeV and also a
non-resonant single-pion background.  The nucleon, $N-\Delta$ and $N-N^*$ vector form
factors are based on recent electron-scattering data while the axial couplings are
obtained assuming PCAC (partial conservation of the axial current). The $Q^2$ dependence
of the nucleon and $N-\Delta$ axial form factors, $F_A$ and $C_5^A$, is fitted to bubble
chamber neutrino-scattering data.

The neutrino-nucleon cross sections are modified in the nuclear medium. Bound nucleons are
treated within a local Thomas-Fermi approximation. They are exposed to a mean-field
potential which depends on density and momentum. We account for this by evaluating the
above cross sections with full in-medium kinematics, i.e., hadronic tensor and phase-space
factors are evaluated with in-medium four-vectors. We also take Pauli blocking and
collisional broadening of the outgoing hadrons into account. Our model for
neutrino-(bound)nucleon scattering is described in detail in \refcite{Leitner:2008ue}.

After the initial neutrino-nucleon interaction, the produced particles propagate out of
the nucleus. During propagation they undergo FSI which are simulated with the
coupled-channel semi-classical GiBUU transport model (for details, see \refcite{gibuu} and
references therein).  It is based on the BUU equation which describes the space-time
evolution of a many-particle system in a mean-field potential including a collision term.
Nucleons and resonances acquire medium-modified spectral functions and are propagated
off-shell. Herby we ensure, that vacuum spectral functions are recovered after leaving the
nucleus.  The collision term accounts for changes (gain and loss) in the phase-space
density due to elastic and inelastic collisions between the particles, and also to
particle decays into other hadrons. Baryon-meson two-body interactions (e.g., $\pi N \to
\pi N$) are described by resonance contributions and a small non-resonant background term;
baryon-baryon cross sections (e.g., $NN \to NN$, $R N \to N N$, $R N \to R' N$, $N N \to
\pi NN$) are either fitted to data or calculated, e.g., in pion exchange models. The
three-body channels $\pi N N \to NN$ and $\Delta N N \to NNN$ are also included. The BUU
equations for all particle species are thus coupled through the collision term and also
through the potentials. Such a coupled-channel treatment is required to account for side
feeding into different channels.

\section{CCQE and CC$1\pi^+$ entanglement}

One challenge is to identify \emph{true} CCQE events in the detector, i.e., muons
originating from an initial QE process. To be more precise, true CCQE corresponds to the
inclusive CCQE cross section including all medium effects, or, in other words, the CCQE
cross section before FSI.  The difficulty comes from the fact that the true CCQE events
are masked by FSI in a detector built from nuclei.  The FSI lead to misidentified events,
e.g., an initial $\Delta$ whose decay pion is absorbed or which undergoes ``pion-less
decay'' can count as CCQE event --- we call this type of background events ``fake CCQE''
events. We denote every event which looks like a CCQE event by ``CCQE-like''.

In general, at Cherenkov detectors such as MiniBooNE CCQE-like events are all those where
no pion is detected while in tracking detectors such as K2K-SciBar/SciFi CCQE-like events
are those where a single proton track is visible and at the same time no pions are
detected.

\begin{figure}[tbp]
  \includegraphics[scale=\plotscale]{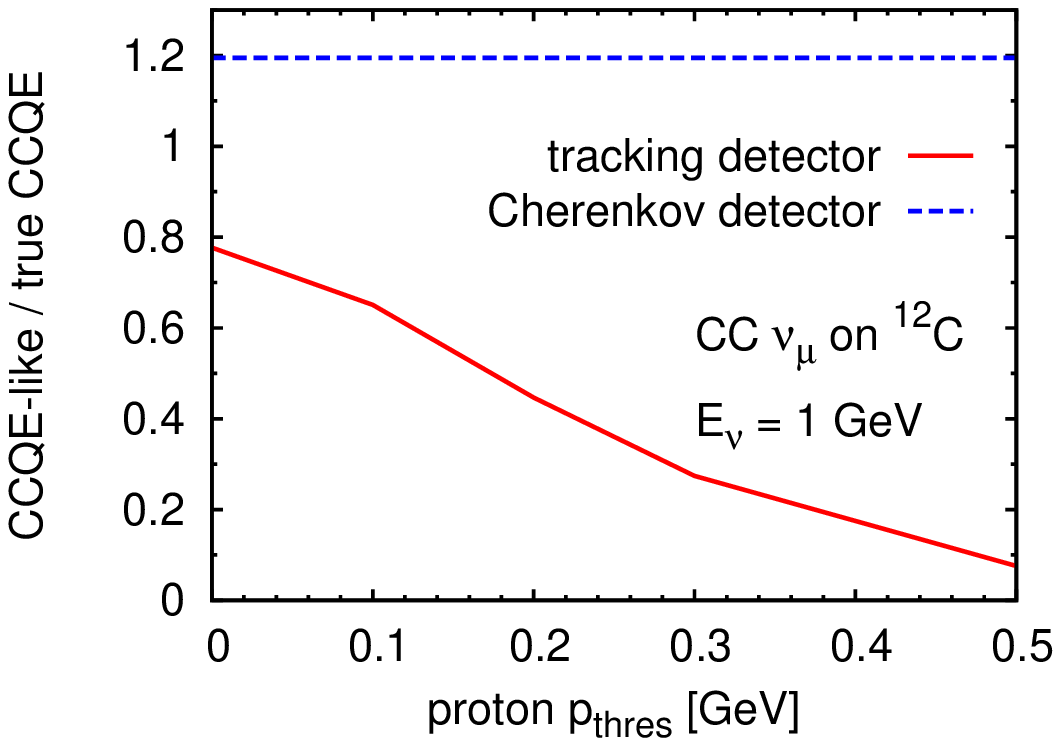}
  \includegraphics[scale=\plotscale]{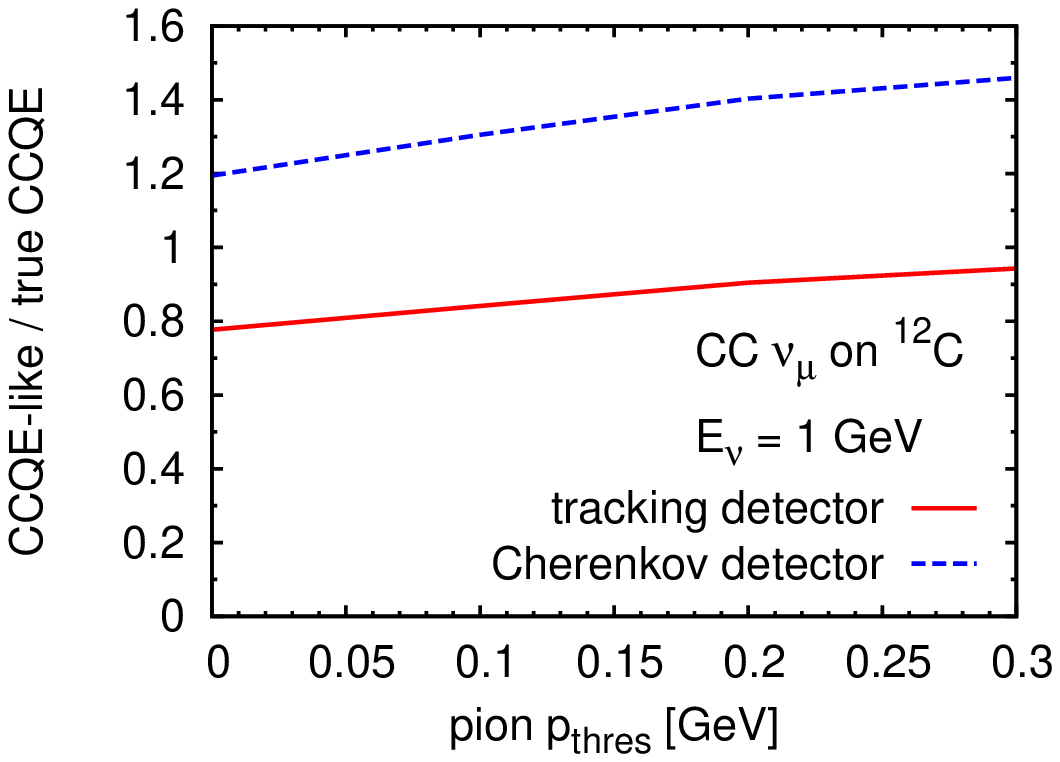}
    \caption{Ratio of the CCQE-like to the true CCQE cross section as a function of
  the proton (pion) momentum detection threshold for CC $\nu_\mu$ on \carbon{} at
  $E_\nu=1$ GeV. The solid lines are obtained using the tracking detector identification,
  while the dashed lines are for Cherenkov detectors.
  \label{fig:CCQElike_over_trueCCQE_thresholds}}
\end{figure}

To investigate the relation between the CCQE-like and true CCQE cross section, we show
their ratio as a function of proton and pion momentum thresholds in
\reffig{fig:CCQElike_over_trueCCQE_thresholds}. As the proton is not at all relevant for
the CCQE identification in Cherenkov detectors, the ratio is independent of the proton
momentum detection threshold (dashed line in left panel). This is very different in
tracking detectors which rely on the detected proton --- here the efficiency is reduced to
$\approx$10\% at a proton momentum threshold of 0.5 GeV (solid line in left panel). Even
at $|\vec{p}|_\text{thres}^p=0$ the efficiency does not exceed 80\% because of
charge-exchange processes that lead to the emission of undetected neutrons and because of
secondary proton knockout that leads to multiple-proton tracks.  Focussing on the right
panel of \reffig{fig:CCQElike_over_trueCCQE_thresholds} we find that the CCQE-like cross
section increases for both detector types as $|\vec{p}|_\text{thres}^\pi$ increases. In
this case even more events with pions in the final state appear as CCQE-like because these
pions are below threshold and thus not detected.

Fixing the flux normalization with HARP's pion-production data, the MiniBooNE
collaboration has presented their first, preliminary absolutely normalized total,
differential and double differential cross sections for CCQE at this conference and finds
an excess of about 35\% compared to the total cross section measured at NOMAD, ANL and BNL
\cite{teppeiKatoriNUINT09Talk}.  We strongly emphasize that these absolute cross sections
depend directly on the pion background subtraction which again is based on the Monte Carlo
prediction (cf., \refcite{teppeiKatoriNUINT09Talk}).

\begin{figure}[tbp]
  \includegraphics[scale=\plotscale]{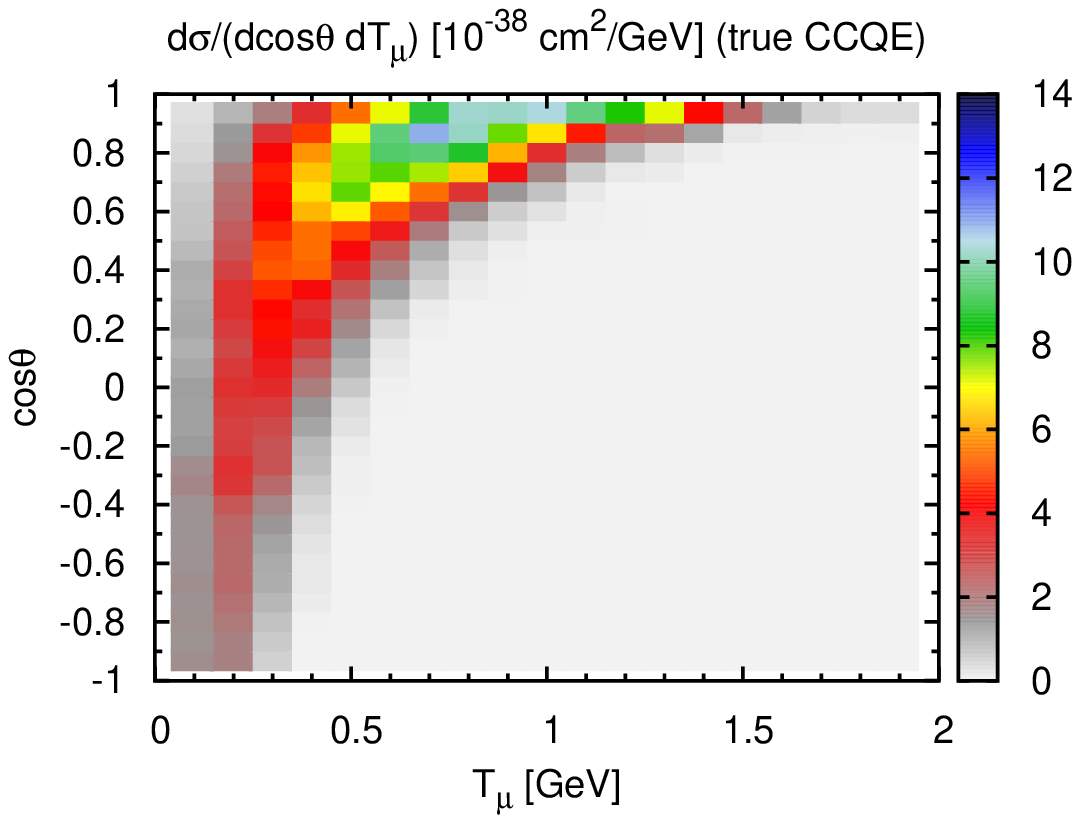}
  \includegraphics[scale=\plotscale]{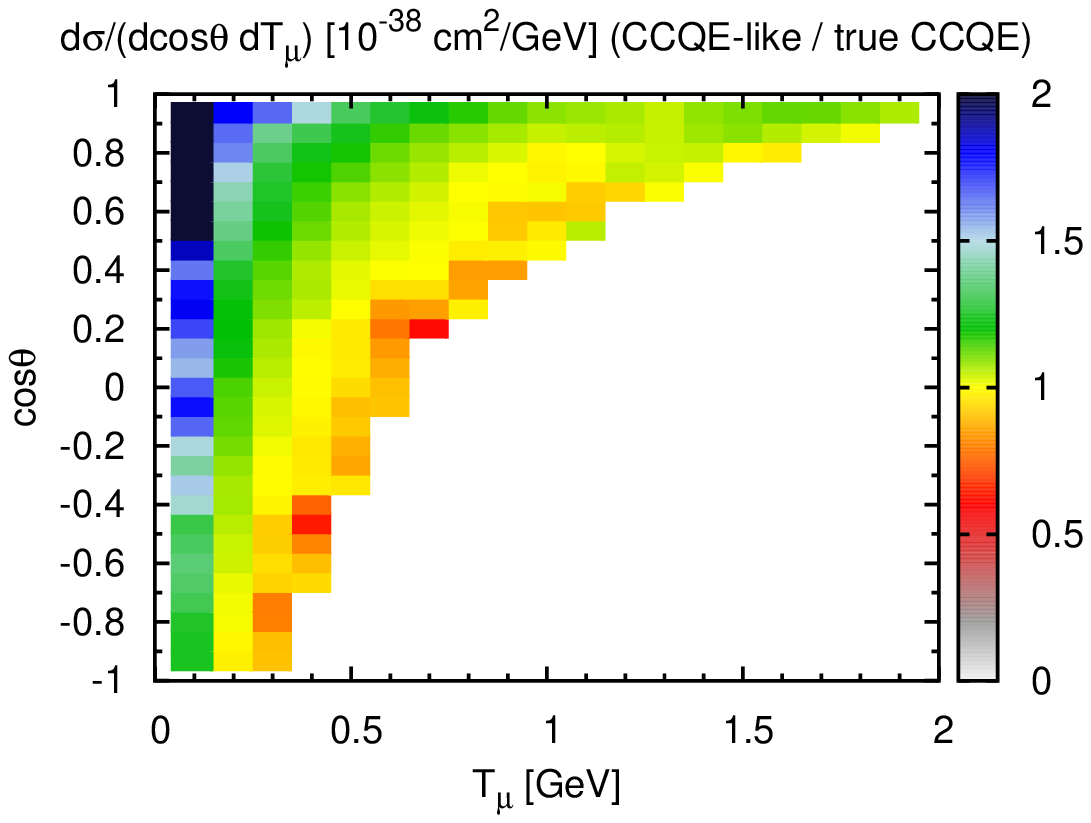}
    \caption{Double differential cross section on \carbon{} averaged over the
  MiniBooNE flux as a function of the muon kinetic energy and the muon scattering angle.
  The left panel shows the true CCQE cross section, the right panel the ratio of the CCQE-like to the true CCQE cross section. 
  \label{fig:MiniBooNE_doublediff_QE}}
\end{figure}

In \reffig{fig:MiniBooNE_doublediff_QE}, we show our prediction for the double differential
cross section at MiniBooNE in muon observables, all calculated with $M_A=1$ GeV. The left
panel shows the true CCQE events. To emphasize the role of ``fake'' CCQE events, we show
the ratio CCQE-like/true CCQE in the right panel. Unlike for monochromatic beams, the QE and
$\Delta$ peaks are not distinguishable any more but strongly overlap. This fact makes a
model-independent cut based on muon variables to subtract the background impossible.

\section{MiniBooNE's CC1$\pi^+$ measurement}
\begin{figure}[tbp]
  \includegraphics[scale=\plotscale]{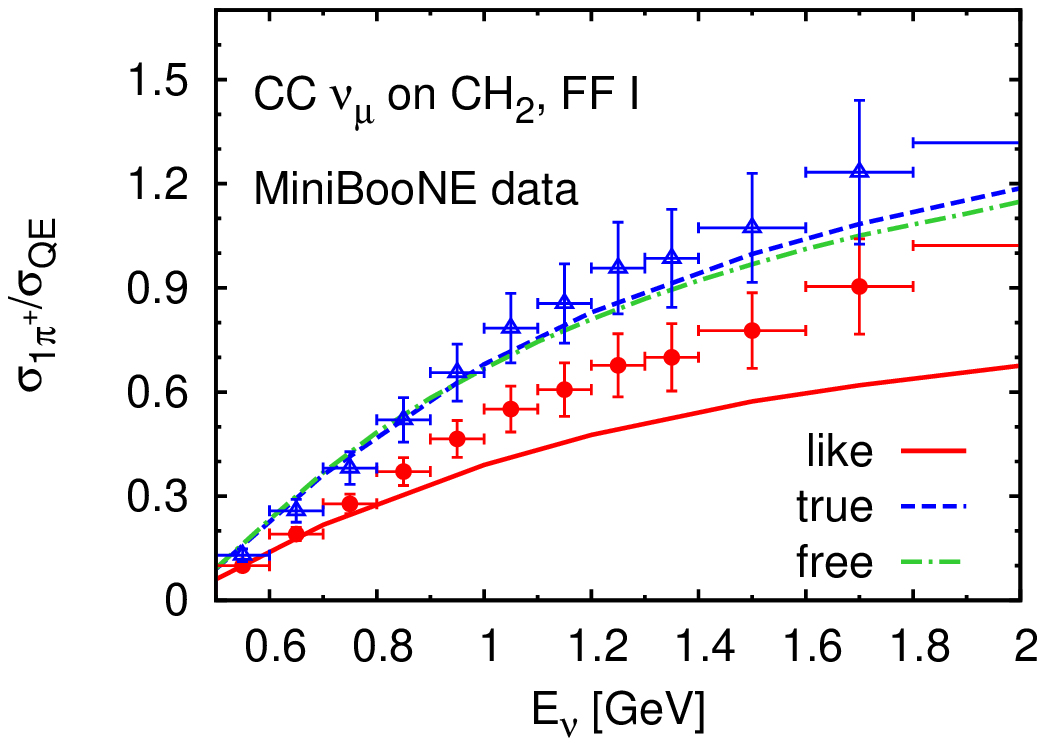}
  \includegraphics[scale=\plotscale]{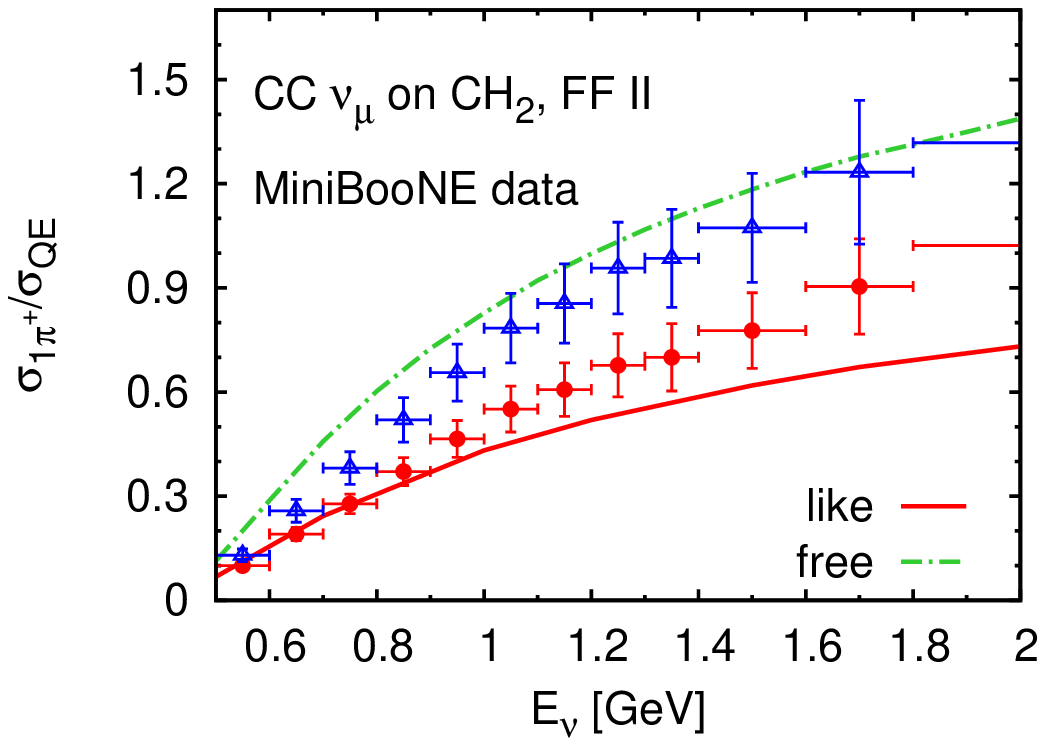}
  \caption{Left panel: single-$\pi^+$/QE cross section ratio for CC interactions
    vs.~neutrino energy on CH$_2$ together with recent data from MiniBooNE
    \cite{AguilarArevalo:2009eb} (upper data set: corrected for FSI, lower data set: uncorrected for
    FSI). The solid lines denote the
    CC$1\pi^+$-like/CCQE-like result (Cherenkov detector definitions), the dashed lines
    stand for the true CC$1\pi^+$/true CCQE result, and the dash-dotted lines give the
    vacuum expectation, i.e., the sum of the nucleon cross sections (with two additional
    protons in the MiniBooNE case).  Right panel: same but with a dipole form for $C_5^A$ (see text).
    \label{fig:MB_CC}}
\end{figure}
In the left panel of \reffig{fig:MB_CC} we give our results for the single-$\pi^+$/QE
ratio for CC interactions on mineral oil CH$_2$. The solid lines denote the
CC$1\pi^+$-like/CCQE-like result, the dashed lines stand for the true CC$1\pi^+$/true CCQE
result, and the dash-dotted lines give the vacuum expectation. Note that we have applied
the Cherenkov detector identification criteria.

We emphasize that nuclear corrections cancel out in the ratio, only as long as FSI are not
considered (``true'' vs.~``free''). In general, the complexity of FSI prevent such
cancellations as one can infer from the result denoted with ``like'' which does not coincide with
the ``true'' and ``free'' ones.

We further compare to very recent MiniBooNE data \cite{AguilarArevalo:2009eb}
(\reffig{fig:MB_CC} left). Let us first focus on the data denoted with the
triangles (upper data set). These are corrected for FSI using a specific Monte Carlo
generator, i.e., they give the cross sections for bound nucleons ``before FSI''. As this
procedure introduces a model dependence in the data, a fully consistent comparison is not
possible.  Ignoring this inconsistency, our calculation denoted with ``true'' should be the
one to compare with.  The agreement is perfect for energies up to 1.5 GeV, and still
within their error bars for higher $E_\nu$. The MiniBooNE data denoted with bullets (lower
data set) is their result for the ratio of CC$1\pi^+$-like to CCQE-like. As these data are
not corrected for FSI within a specific Monte Carlo event generator, this observable is
less model dependent. Still, the energy reconstruction requires specific assumptions as
well as the detector simulation.  We find that our calculation clearly underestimates the
uncorrected data. However, the perfect agreement with their corrected distribution
indicates a significant difference between the pion absorption models.

\begin{figure}[tbp]
  \includegraphics[scale=\plotscale]{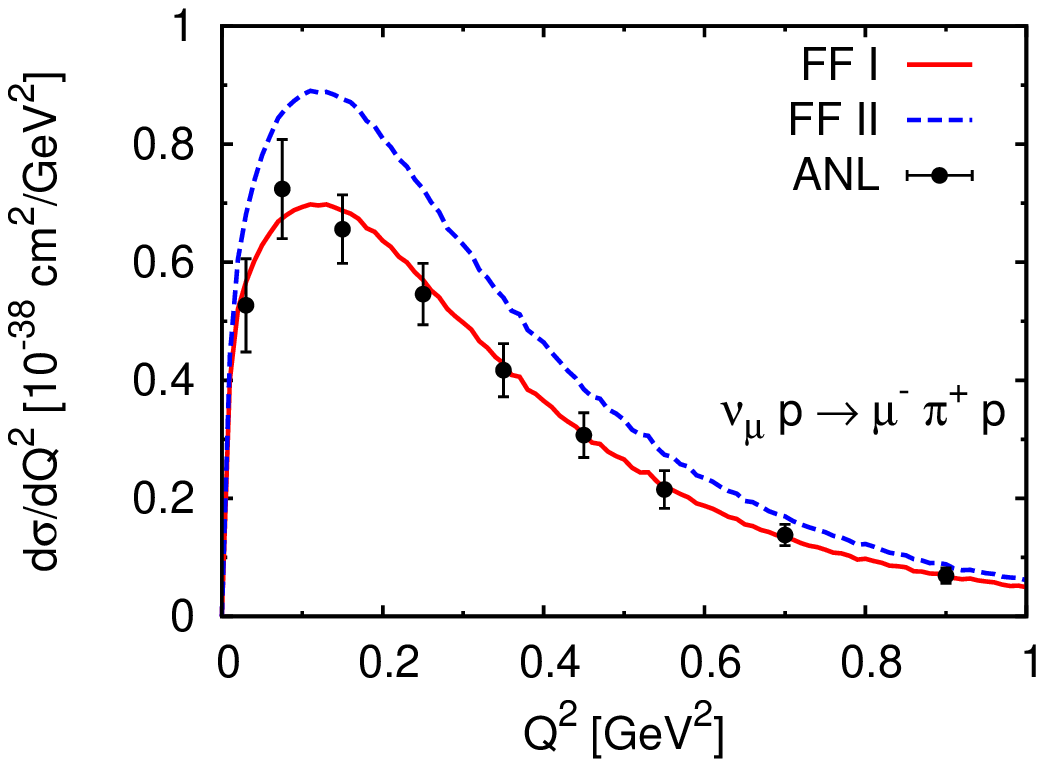}
  \includegraphics[scale=\plotscale]{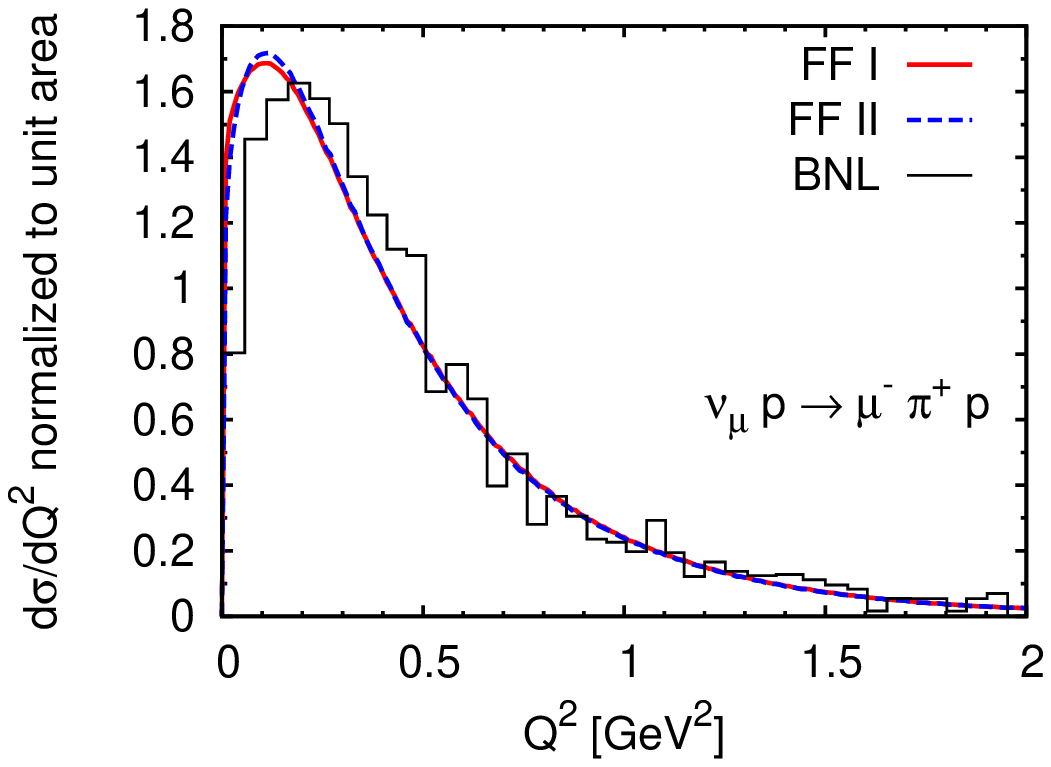}
  \caption{Differential cross section d$\sigma/$d$Q^2$ averaged over the ANL flux (left
    panel) and the BNL flux (right panel) [invariant mass cut at $W<1.4$ GeV]. The data
    are taken from \refscite{Radecky:1981fn,Kitagaki:1986ct}. The form factors are
    detailed in the text.
    \label{fig:nucleon_Qs_CC}}
\end{figure}

\begin{figure}
  \includegraphics[scale=\plotscale]{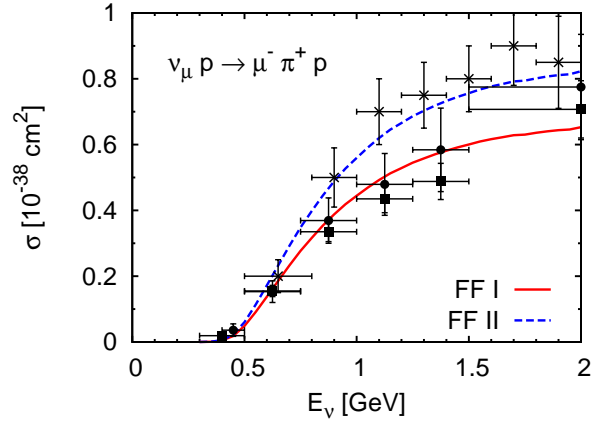}
  \caption{Total CC pion production cross sections compared to the pion production data of
    ANL [Refs.~\cite{Barish:1978pj} ($\bullet$), \cite{Radecky:1981fn} ($\blacksquare$)]
    and BNL [\cite{Kitagaki:1986ct} ($\times$)] The solid line has been obtained with
    $\left[C_5^A(Q^2)\right]_I$ while $\left[C_5^A(Q^2)\right]_{II}$ was used for the
    dashed line.
    \label{fig:nucleon_sigtot_CC}}
\end{figure}

The underestimate of the pion/quasielastic ratio in particular at higher energies could be
due to, among other possibilities, an underestimate of the pion production cross section
or an overestimate of the CCQE-like cross section. Both, which means for the latter the
fake CCQE events, depend directly on the input at the nucleon level, i.e., in particular
on the axial form factor $C_5^A$ of the $\Delta$ resonance (see \refcite{Leitner:2008ue}
for details).  In the above calculations, we have used
\begin{equation}
  \left[C_5^A(Q^2)\right]_I = {{C_5^A(0)\left[ 1+\frac{a Q^2}{b+Q^2} \right] } {\left( 1+ \frac{Q^2}{{M_A^I}^2}\right)^{-2}}} ; \label{eq:axialformfactorform}
\end{equation}
with $a=-0.25$, $b=0.04$ GeV$^2$, $M_A^I=0.95$ GeV obtained from a fit to the ANL bubble
chamber data and the PCAC value of $C_5^{A}(0)=1.17$.  However, the ANL and BNL results
differ with the BNL result being approximately 30\% higher. To account for the BNL
findings, we introduce --- following \refcite{Graczyk:2009bc} --- a dipole form factor
\begin{equation}
  \left[C_5^A(Q^2)\right]_{II} = C_5^A(0) {\left( 1+ \frac{Q^2}{{M_A^{II}}^2}\right)^{-2}} ,
\end{equation}
with $M_A^{II}=0.94$ GeV. The results for both parametrizations are shown in
\reffig{fig:nucleon_Qs_CC} and \reffig{fig:nucleon_sigtot_CC} together with the data of
ANL and BNL.  The shape of the BNL distribution is described well by both form factors
(right panel of \reffig{fig:nucleon_Qs_CC}) while the ANL data are overestimated by the
dipole form. Still one should bear in mind that there is a small non-resonant background
\cite{Hernandez:2007qq} that has been neglected in this channel.  For a discussion of
possible Deuterium effects, in particular at low $Q^2$, see
\refscite{Graczyk:2009bc,Alvarez-Ruso:1998hi}.

In the right panel of \reffig{fig:MB_CC} we present our results for the single-$\pi^+$/QE
ratio using the dipole form factor. We find a slight enhancement of the ratio but still
underestimate the data. Note that the enhancement is much more moderate than in
\reffig{fig:nucleon_sigtot_CC}.  This is due to the fact, that both, numerator and
denominator, i.e., also the CCQE-like cross section, are increased when the dipole form
factor is used and, as a consequence, the larger CCQE-like cross section compensates the
enhancement in the pion cross section.  From this comparison we conclude that an increase
of the total pion production cross section on the nucleon compatible with the BNL data
seems to be insufficient to describe this ratio at all energies. A similar result for the
ratio has been recently obtained by Athar \refetal{Athar:2009rc}.

\section{NC$1\pi^0$}

\begin{figure}
  \includegraphics[scale=\plotscale]{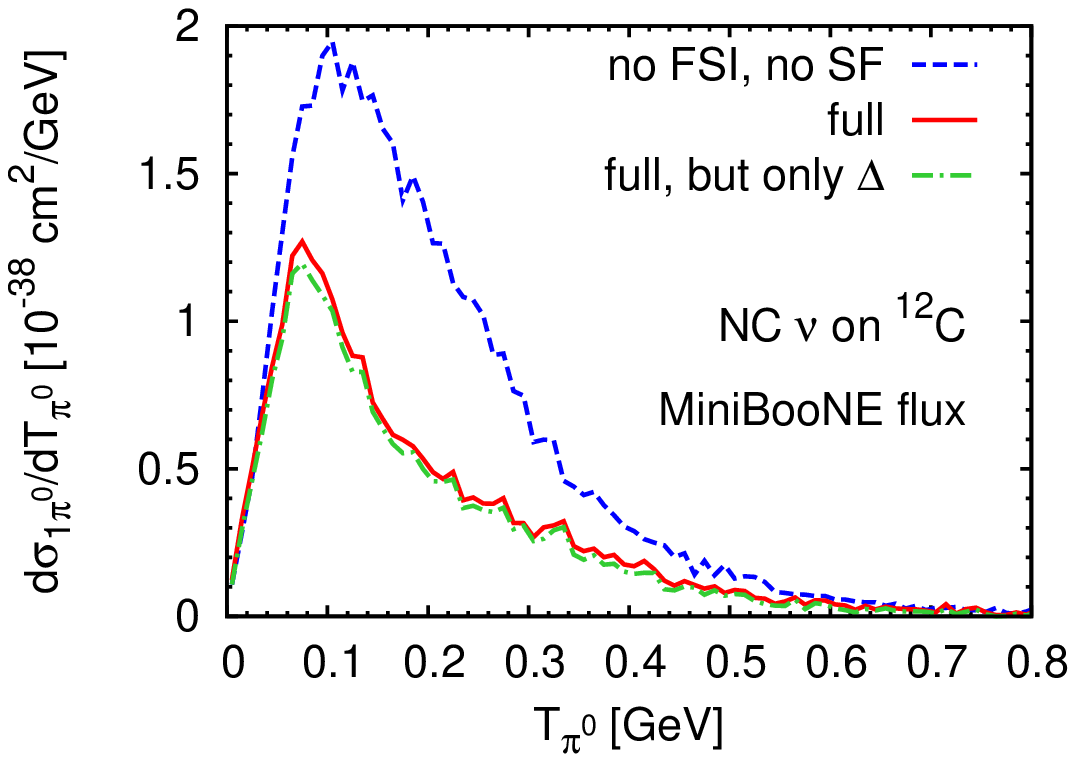}
  \includegraphics[scale=\plotscale]{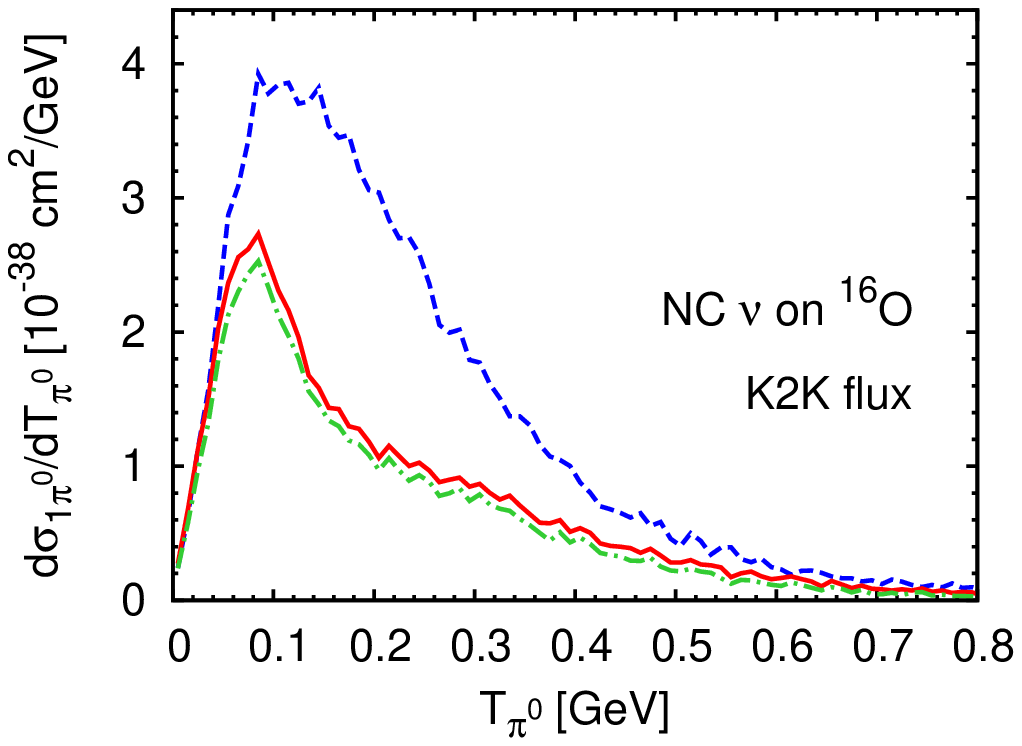}
  \caption{Left panel: NC
  induced single-$\pi^0$ production on \carbon{} as a function of the pion kinetic energy
  averaged over the MiniBooNE flux. The results were obtained with $\left[C_5^A(Q^2)\right]_I$. Right panel: same on \oxygen{} averaged over the K2K
  flux. The dashed lines show our calculation without FSI or spectral functions, both
  included in the full calculation denoted with the solid lines. The dash-dotted lines
  indicate the contribution from the $\Delta$ resonance to the full calculation.
  \label{fig:MiniBooNE_K2K_NC}}
\end{figure}

In the left panel of \reffig{fig:MiniBooNE_K2K_NC}, we show our results for NC
single-$\pi^0$ production off \carbon{} as a function of the pion kinetic energy.  We have
averaged over the MiniBooNE energy flux which peaks at about 0.7 GeV neutrino energy
\cite{AguilarArevalo:2008yp}.  In NC reactions the total pion yield is dominated by
$\pi^0$ production, while $\pi^+$ dominate in CC processes (for details, see
\refcite{Leitner:2008wx} and references therein).  Comparing the dashed with the solid
line (results without FSI and spectral function vs.~full calculation), one finds a
considerably change. The shape is caused by the energy dependence of the pion absorption
and rescattering cross sections.  Pions are mainly absorbed via the $\Delta$ resonance,
i.e, through $\pi N \to \Delta$ followed by $\Delta N \to NN$. This explains the reduction
in the region around $T_\pi=0.1-0.3$ GeV.  Pion elastic scattering $\pi N \to \pi N$
reshuffles the pions to lower momenta and leads also to charge exchange scattering into
the charged pion channels.  The vast majority of the pions comes from initial $\Delta$
excitation (dash-dotted line), their production in the rescattering of nucleons is not
significant at these energies.

The right panel of \reffig{fig:MiniBooNE_K2K_NC} shows the
results for NC single-$\pi^0$ production off \oxygen{} averaged over the K2K energy flux
which peaks at about 1.2 GeV neutrino energy \cite{Nakayama:2004dp}. Compared to the left panel, the spectrum is
broader and extends to larger $T_\pi$ due to the higher neutrino energy. The reduction in
the region around $T_\pi=0.1-0.3$ GeV is mainly caused by the pion absorption via the
$\Delta$ resonance (compare dashed and solid lines). Again, pion production through
initial QE scattering is not sizable.

\section{Final remark}
Overall, the impact of FSI on observables is dramatic and not at all negligible.  A
qualitatively and quantitatively correct treatment is thus of great importance.  More
model-independent data are certainly needed.

\section{Acknowledgments} 
We thank O. Lalakulich for fruitful discussions. This work has
been supported by the Deutsche Forschungsgemeinschaft.


\end{document}